\documentclass[twocolumn,nofootinbib]{revtex4-1}

\usepackage{amsmath}
\usepackage{amsfonts}
\usepackage{amssymb}
\usepackage{graphicx}
\usepackage{pifont}

\usepackage{color}
\usepackage{verbatim}

\usepackage{simplemargins}
\setbottommargin{2.6cm}

\newcommand{\D}{\mathrm{d}}
\newcommand{\acos}{\mathrm{arccos}}

\begin{document}


\title{Students' confusions about the electric field of a uniformly moving charge}

\author{Petar \v{Z}ugec}
\email{pzugec@phy.hr, pzugec.phy@pmf.hr}
\affiliation{Department of Physics, Faculty of Science, University of Zagreb, Zagreb, Croatia}

\author{Davor Horvati\'{c}}
\affiliation{Department of Physics, Faculty of Science, University of Zagreb, Zagreb, Croatia}

\author{Ivica Smoli\'{c}}
\affiliation{Department of Physics, Faculty of Science, University of Zagreb, Zagreb, Croatia}



\begin{abstract}

In light of a recent direct experimental confirmation of a Lorentz contraction of Coulomb field (an electric field of a point charge in a uniform motion), we revisit some common confusions related to it, to be mindful of in teaching the subject. These include the questions about a radial nature of the field, a role of the retardation effect due to a finite speed of information transfer and some issues related to a depiction of Coulomb field by means of the Lorentz contracted field lines.

\end{abstract}

\maketitle

\section{Introduction}

Relativistic transformation of an electric field---particularly that of a point charge in a uniform motion (a Coulomb field)---is a regular topic of (under)graduate physics courses such as General Physics and dedicated courses on Classical Electrodynamics. In light of a recent groundbreaking \textit{direct} experimental confirmation of a Lorentz contraction of Coulomb field \cite{nature}, we expect that this subject will become even more prominent in teaching the special relativistic aspects of Classical Electrodynamics. We use this opportunity to address some seemingly widespread confusions (if not misconceptions) that are often encountered in teaching these. We have indeed observed these confusions in a classroom setting. Through an active interaction with the generations of students we have had an opportunity to develop an effective discussion, presented herein. Some references to these confusions---mostly unfocused or mentioned in passing---may be found scattered throughout literature~\cite{feynman,griffiths,purcell}. Explicitly exposing students (in a well-timed manner) to questions raised herein should certainly help them in developing a deeper understating and a well deserved appreciation for the internal consistency of both Classical Electrodynamics and Special Relativity. In that, a close connection between a charge motion and the Special Relativity itself is aptly reflected by a very title of Einstein's foundational paper `On the Electrodynamics of Moving Bodies'~\cite{einstein}.

The goal of our discussion is twofold: (1)~to introduce the confusing issues in a compelling and meaningful way, so as to arouse the students' interest in what may otherwise be seen as a dry subject; (2)~and to resolve these issues in an insightful way that should yield a greater appreciation of a deep meaning behind seemingly uninformative equations. At the same time, bringing these issues to the explicit attention of teachers should help \textit{them} in recognizing and appreciating the students' struggles with the basic facts about the dynamic electric fields. In turn, this should help them in properly addressing these issues before students' struggles get out of hand.

\section{First things first}


In this work we wish to keep things as simple as possible, for maximum clarity. For this reason we consider only the electric field of a point charge in a uniform motion (a Coulomb field), entirely excluding all magnetic fields from a discussion\footnote{
We would, of course, have to account for a magnetic field of a moving charge if we were interested in any specific effect upon other charges. But in this work we are only interested in an electric field itself.
}. Furthermore, whenever a decomposition of vectors into components is required, we will use two-dimensional coordinate system. In that, the charge velocity $\mathbf{v}$ will be directed along the $x$-axis (\mbox{$\mathbf{v}=v\hat{\mathbf{x}}$}), so that the $x$-component $E_x$ of the electric field \mbox{$\mathbf{E}=E_x\hat{\mathbf{x}}+E_y\hat{\mathbf{y}}$} is understood to be parallel to the charge motion, while the $y$-component $E_y$ is understood to be perpendicular to it\footnote{
In order to immediately achieve a correct three-dimensional description, one only needs to replace the Cartesian $x$ and $y$ coordinates with the cylindrical $z$ and $\rho$ coordinates (\mbox{$x\to z$} and \mbox{$y\to \rho$}), so that \mbox{$\mathbf{E}=E_z\hat{\mathbf{z}}+E_\rho\hat{\boldsymbol{\rho}}$}.
}. In this sense, let $\mathbf{E}$ be a \textit{purely electrostatic} field of a point charge at rest, in a frame where no magnetic fields are present (\mbox{$\mathbf{B}=\mathbf{0}$}). The Lorentz transformed components of a Coulomb field $\mathbf{E}'$ in a frame where the charge moves with the velocity \mbox{$\mathbf{v}=v\hat{\mathbf{x}}$} are then:
\begin{align}
&E'_x=E_x,
\label{lor_ex}\\
&E'_y=\gamma E_y,
\label{lor_ey}
\end{align}
with \mbox{$\gamma=(1-v^2/c^2)^{-1/2}$} as a standard Lorentz factor. These tranformations are, of course, only `remainders' of the full field transformations which also take into account a presence of the magnetic field. The listing of these transformations---and different forms of writing them out---may be found in any standard textbook. 

\begin{figure*}[t!]
\centering
\includegraphics[width=0.85\textwidth,keepaspectratio]{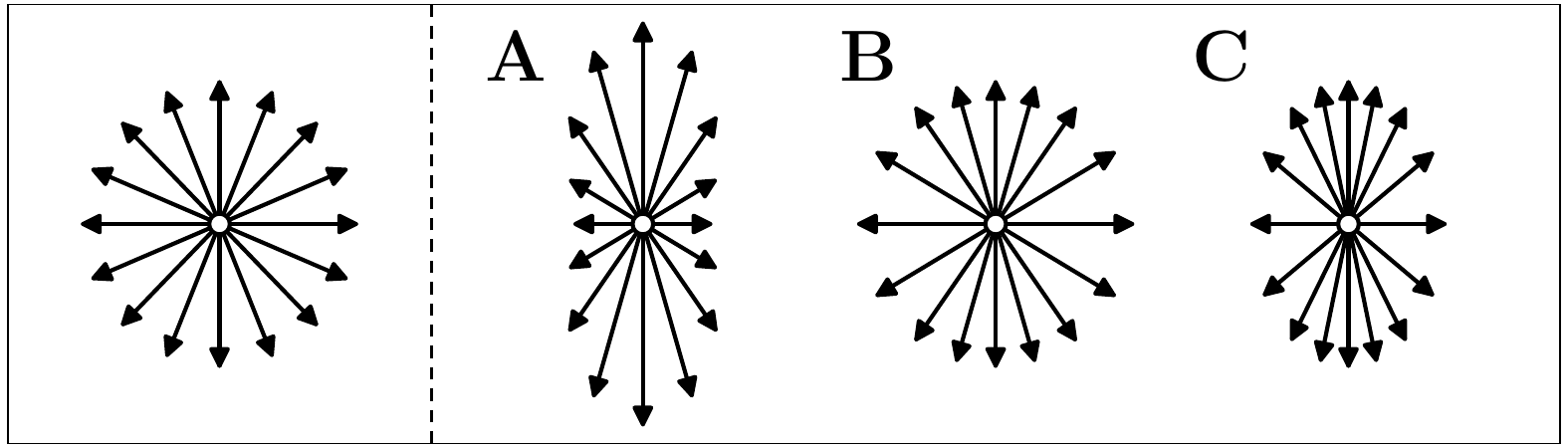}
\caption{Field lines depicting a Lorentz contraction of a Coulomb field of a positive point charge. Leftmost configuration of field lines shows an electric field of a charge at rest. Configurations A, B, C (attempt to) show a field of a charge in a uniform horizontal motion (either to the left or right) with speed \mbox{$v=0.7c$} (\mbox{$\gamma\approx 1.4$}). These depictions are addressed in Section~\ref{field_lines}.
} 
\label{fig1}
\end{figure*}

Thus transformed field is often represented by means of `Lorentz contracted' field lines (quotations intentional), and usually by one of the depictions from figure~\ref{fig1}. Leftmost (`isotropic') configuration of field lines represents en electric field around a point charge at rest. `Lorentz contracted' configurations A, B, C represent a Lorentz transformed Coulomb field of a point charge in a uniform motion. For example, one can find the second~(B) representation in a form of figure~5.15 from~\cite{purcell} and the third~(C) representation in a form of figure~26-4 from~\cite{feynman}. While the first~(A) representation is most seldomly used, we claim that it is the most appropriate one (as far as using the field lines for representing a vector field is appropriate at all). There is a rich discussion to be had on this issue. We postpone it until Section~\ref{field_lines}, when we will have established sufficient grounds for a discussion. For now we only require the visual suggestions from any of these displays, as they present themselves throughout the literature. 

\section{Radial or not?}
\label{sec_radial}

Every student of physical sciences knows (or should know) that the electric field of a point charge at rest is radial. Let $\mathbf{r}_0$ be the instantaneous position of a point charge. Let $\mathbf{r}$ be an arbitrary point in space where we observe its electric field. Introducing \mbox{$\mathbf{R}= \mathbf{r}-\mathbf{r}_0=R_x\,\hat{\mathbf{x}} +R_y\,\hat{\mathbf{y}}$} as a charge-relative position and \mbox{$\hat{\mathbf{R}}=\mathbf{R}/|\mathbf{R}|$} as its unit vector, this basic fact may be expressed as:
\begin{equation}
\mathbf{E}(\mathbf{r})=E \,\hat{\mathbf{R}}=E\,\frac{R_x\, \hat{\mathbf{x}}+R_y\, \hat{\mathbf{y}}}{\sqrt{R_x^2+R_y^2}}=E_x\,\hat{\mathbf{x}}+E_y\,\hat{\mathbf{y}}.
\end{equation}
The radial nature of the field lies in the fact that the field components $E_x$ and $E_y$ are related precisely as the components of vector $\mathbf{R}$:
\begin{equation}
\frac{E_x}{E_y}=\frac{R_x}{R_y}.
\label{radial}
\end{equation}

Let us now think from a perspective of a student who has either never seen a \textit{transparent} derivation of the electric field transformations or has long since forgotten it (which is why the students in later years are especially susceptible to the forthcoming confusion). Left only with the end results (\ref{lor_ex}) and (\ref{lor_ey})---which are easily found in any textbook and are usually written precisely in this form---a student might be led to a following conclusion. If we observe the relation between the Lorentz transformed field components:
\begin{equation}
\frac{E_x'}{E_y'}=\frac{1}{\gamma}\frac{E_x}{E_y}=\frac{1}{\gamma}\frac{R_x}{R_y},
\label{Rxy_old}
\end{equation}
it seems that the transformed field should no longer be radial; if the ratio $E_x/E_y$ is `radial', then the ratio $E'_x/E'_y$ deviates from the `radial one' by factor $1/\gamma$. But wait! Why do the plots like those from figure~\ref{fig1} still depict the Coulomb field of a moving charge as radial?

We urge the reader to take a moment to appreciate this confusion, as it may appear from a student's perspective. The answer, of course, is that the transformed field is radial in a \textit{new} frame, wherein the spacetime coordinates are also Lorentz transformed. Thus, the components of the transformed field are supposed to satisfy:
\begin{equation}
\frac{E'_x}{E'_y}=\frac{R'_x}{R'_y},
\label{Rxy_new}
\end{equation}
where all quantities are from the \textit{same} frame. In order to confirm that this is the case, one just needs to observe the Lorentz transformations of the spatial lengths $R_x$ and $R_y$. Since $R_y$ is in the direction perpendicular to the charge motion---where we now consider a charge velocity as a velocity between the observer frames---it remains unchanged: \mbox{$R'_y=R_y$}. On the other hand, since the transformed field \mbox{$\mathbf{E}'(\mathbf{r}')=E' \,\hat{\mathbf{R}}'$} is determined by the charge-relative position \mbox{$\mathbf{R}'= \mathbf{r}'-\mathbf{r}'_0$} between the points $\mathbf{r}'$ and $\mathbf{r}'_0$ \textit{simultaneous in a new frame}, a length contraction for $R_x$ in a direction of a frame boost immediately applies\footnote{
There is a small subtlety to be appreciated in making a detailed derivation from Lorentz transformations, related to a simultaneity of events in different observer frames. See \mbox{Appendix~\ref{appendix_contraction}}.
}: \mbox{$R_x'=R_x/\gamma$}. Thus having verified the equivalence \mbox{$R_x/\gamma R_y=R'_x/R'_y$} between~(\ref{Rxy_old}) and~(\ref{Rxy_new}), the radial nature of the transformed field is confirmed.

At this point a student might have rediscovered---or even discovered for the first time---that the `Lorentz transformation of electric field' also entails a transformation of spacetime coordinates, aside from a transformation of a field itself. This necessary and unavoidable (rather than coincidental and unexpected) fact may be somewhat obscured when the field transformations are derived by `taking a long road', however correct, via Li\'{e}nard-Wiechert potentials~\cite{feynman,griffiths} or by some other involved method~\cite{jefimenko_direct}. Fortunately, there is a simple and transparent derivation of the field transformations~(\ref{lor_ex}) and~(\ref{lor_ey}), relying precisely on the Lorentz transformation of the coordinates (see, for example, Chapter 12.3.2 from~\cite{griffiths},  or Chapters~5.5 and~5.6 from~\cite{purcell}). The discussion we have presented is a way for a student to (re)discover this `in reverse', after first being led astray by~(\ref{lor_ex}) and~(\ref{lor_ey}).

In the end, entire discussion is nicely rounded by reminding ourselves and students that there is a well known expression for a Coulomb field of a point charge in a uniform motion:
\begin{equation}
\mathbf{E}'(\mathbf{r}')=\frac{q}{4\pi\epsilon_0}\frac{1-v^2/c^2}{[1-(v^2/c^2)\sin^2\theta']^{3/2}}\frac{\hat{\mathbf{R}}'}{|\mathbf{R}'|^2}
\label{Er}
\end{equation}
from which its radial nature is clearly visible. As usual, $q$ is the value of the charge, $\epsilon_0$ the permittivity of vacuum, $c$ the speed of light in vacuum. Angle $\theta'$ is that between~$\mathbf{R}'$ and the charge velocity~$\mathbf{v}$, i.e. \mbox{$\cos\theta'=\hat{\mathbf{R}}'\cdot \hat{\mathbf{v}}$}. For our selection of coordinates: \mbox{$\sin\theta'=R'_y/\big[(R'_x)^2+(R'_y)^2\big]\,\!^{1/2}$}.

\subsection*{Let us put it to the test}

Expression~(\ref{Er}) also presents a perfect opportunity for a student to develop a deep `internalization' of coordinate transformations and their role in field transformations. To this end, let us observe what~(\ref{Er}) yields for \mbox{$\theta'=0$} (a field at a distance $R'_x$ right ahead of a charge):
\begin{equation}
\mathbf{E}'(\theta'=0)=\frac{1}{\gamma^2}\left[\frac{q}{4\pi\epsilon_0}\frac{1}{(R'_x)^2}\right]\hat{\mathbf{x}}
\label{E_para}
\end{equation}
and for \mbox{$\theta'=\pi/2$} (a field at a lateral distance $R'_y$):
\begin{equation}
\mathbf{E}'(\theta'=\pi/2)=\gamma\left[\frac{q}{4\pi\epsilon_0}\frac{1}{(R'_y)^2}\right]\hat{\mathbf{y}}.
\label{E_perp}
\end{equation}
The terms inside the square brackets certainly look like the electrostatic terms, i.e. like the corresponding $x$ and $y$ components of an electric field of a charge at rest. In this sense a pure $E_y$ field from~(\ref{E_perp}) seems to be increased by~$\gamma$, in perfect accordance with the general Lorentz transformation~(\ref{lor_ey}). But wait! A pure $E_x$ field from~(\ref{E_para}) seems to be decreased by \mbox{$1/\gamma^2$}, instead of remaining the same, as in~(\ref{lor_ex}). What gives?

The answer, again, lies in coordinate transformations. The components $E_x$ and $E_y$ from~(\ref{lor_ex}) and~(\ref{lor_ey})---i.e. the components of an electrostatic field \mbox{$\mathbf{E}(\mathbf{r})=(q/4\pi\epsilon_0|\mathbf{R}|^2)\hat{\mathbf{R}}$} of a point charge---take the familiar electrostatic form when expressed in coordinates $R_x$ and $R_y$ from \textit{its rest-frame}:
\begin{align}
&E_x(\theta=0)=\frac{q}{4\pi\epsilon_0}\frac{1}{R_x^2},
\label{E_para0}\\
&E_y(\theta=\pi/2)=\frac{q}{4\pi\epsilon_0}\frac{1}{R_y^2}.
\label{E_perp0}
\end{align}
Since $R'_x$ from~(\ref{E_para}) is Lorentz contracted (\mbox{$R'_x=R_x/\gamma$}), the $E'_x$ component from~(\ref{E_para}) is indeed the same as $E_x$ from~(\ref{E_para0}), in accordance with~(\ref{lor_ex}). Of course, everything is also right with the relation between $E'_y$ and $E_y$, due to a lateral coordinate transforming as \mbox{$R'_y=R_y$}.

This is a crucial point for making of breaking the students' understanding of field transformations~(\ref{lor_ex}) and~(\ref{lor_ey}). Written in that form---as they usually are---they convey a false notion of simplicity. This example shows that by no means would it be redundant to write them with the explicit reference to the appropriate coordinates:
\begin{align}
&E'_x(\mathbf{r}',t')=E_x(\mathbf{r},t),
\label{reminder_x}\\
&E'_y(\mathbf{r}',t')=\gamma E_y(\mathbf{r},t).
\label{reminder_y}
\end{align}
It does not matter that $\mathbf{r}'$ and $t'$ may always be expressed through $\mathbf{r}$ and $t$, and vice versa (supposedly making the distinction of arguments superfluous). This extended form serves as a constant \textit{reminder} that there is much more behind~(\ref{lor_ex}) and~(\ref{lor_ey}) than meets the eye. First and foremost is the fact that the fields from both sides of each equation are the fields \textit{at the same spacetime point}, as measured by different inertial observes. Thus, the field transformations necessarily entail the coordinate transformations. First of all, they bring about a change of a field itself due to a \textit{change of the observer}. But they also (and separately) involve a change of the observer's spacetime coordinates \textit{in which a field is expressed}. In consequence, there may be more factors `popping out' when---starting from a field dependence in one frame, expressed in coordinates from that same frame---we try to express its dependence in some other frame \textit{using the coordinates from a new frame}.

How to clearly illustrate the gist of the matter to a student? While the mathematics itself is clear and simple, the plethora of frame dependent quantities may be distracting, interfering with one's intuitive comprehension, i.e. with `seeing' the issue. In this particular example, a strength of an electric field of a point charge is \textit{in its rest-frame} equal everywhere along the circle around it (remember, we are in 2D; imagine a sphere if you prefer). When the charge is in motion, this equal-strength-circle \textit{from its rest-frame} is Lorentz contracted into equal-strength-ellipse. If we were thinking of the electrostatic components $E_x$ and $E_y$ along the circle, then the Lorentz transformed components $E'_x$ and $E'_y$ lie along the contracted ellipse. If you insist on thinking about the field strength along the circle \textit{in a frame of a moving charge}, you can see that almost all points from this circle are farther from a charge than the points from the contracted equal-strength-ellipse (specifically, they are farther away in the $x$-direction). So the transformed $E'_x$ components along the circle, i.e. at the fixed distance from the moving charge, must be lower than the electrostatic $E_x$ components mapped to the contracted ellipse, due to a weaker field along the circle. (You can also think of a field on this circle around a moving charge as originating from an \textit{elongated} ellipse in its rest frame; the conclusions are the same.) This weakening is the source of $1/\gamma^2$ factor from~(\ref{E_para}). The end result is that among all points at a fixed distance from a point charge in a uniform motion, a ratio between the maximum $E'_y$ and $E'_x$ components is not $\gamma$, as suggested by~(\ref{lor_ex}) and~(\ref{lor_ey}), but rather~$\gamma^3$, as per~(\ref{E_para}) and~(\ref{E_perp}).

\section{Retarded or not?}
\label{sec_retarded}

Let us now approach a case of a uniformly moving point charge from a more general direction: from a well known expression for an electric field of a point charge in an \textit{arbitrary} motion. In the last section we have used primed quantities (such as~$\mathbf{E}'$, $\mathbf{r}'$, $\mathbf{R}'$, $t'$) for a charge in motion, as opposed to a charge at rest. In this section we will only be interested in a moving charge so, for simplicity, we omit the primed notation and use the unprimed one. An electric field of a point charge in arbitrary motion is then~\cite{field1,field2,field3}:
\begin{equation}
\mathbf{E}(\mathbf{r},t)=\frac{q}{4\pi\epsilon_0}\left[\frac{\hat{\mathbf{R}}-\boldsymbol{\beta}}{\gamma^2K^3R^2}+\frac{\hat{\mathbf{R}}\times((\hat{\mathbf{R}}-\boldsymbol{\beta})\times\mathbf{a})}{c^2K^3R}\right]_\tau .
\label{E_general}
\end{equation}
The terms $q$, $\epsilon_0$, $c$ are the same as in~(\ref{Er}). Now unprimed~$\hat{\mathbf{R}}$, $\mathbf{r}$, $t$ are the same as previously primed quantities, with \mbox{$R=|\mathbf{R}|$} as a simplified notation for the vector norm (the full vector \mbox{$\mathbf{R}=\mathbf{r}-\mathbf{r}_0$} being a position of a field-observation point~$\mathbf{r}$ relative to a charge position~$\mathbf{r}_0$). Newly appearing quantities are a typical relativistic abbreviation \mbox{$\boldsymbol{\beta}=\mathbf{v}/c$}, a charge acceleration \mbox{$\mathbf{a}=\D \mathbf{v}/\D t$} and \mbox{$K=1-\hat{\mathbf{R}}\cdot\boldsymbol{\beta}$}. As usual, the Lorentz factor is \mbox{$\gamma=(1-\boldsymbol{\beta}\cdot\boldsymbol{\beta})^{-1/2}$}. Everything within the brackets~$[\cdot]_\tau$ is to be evaluated at the retarded time~$\tau$ such that \mbox{$\tau+R(\tau)/c=t$}, due to a finite time required for the information transfer between~$\mathbf{r}_0$ and~$\mathbf{r}$.

Since we are only interested in a uniformly moving charge (\mbox{$\mathbf{a}=\mathbf{0}$}), we focus on a reduced form of~(\ref{E_general}):
\begin{equation}
\mathbf{E}_{\mathbf{a}=\mathbf{0}}(\mathbf{r},t)=\frac{q}{4\pi\epsilon_0\gamma^2}\left[\frac{\hat{\mathbf{R}}-\boldsymbol{\beta}}{K^3R^2}\right]_\tau .
\label{E_ret}
\end{equation}
This expression for an electric field of a uniformly moving charge consists of two conceptual contributions. The one within the square brackets we will call a relativistic \textit{deformation}. The other one, denoted by brackets $[\cdot]_\tau$ themselves, is a familiar signal \textit{retardation}.

But wait (again)! Have we not already seen some formulas for a charge in a uniform motion---(\ref{lor_ex}) and (\ref{lor_ey}), together with~(\ref{Er})---which make no (explicit) reference to a signal retardation? Are, perhaps, earlier formulas incomplete, accounting only for a field deformation (as we have defined the term), without retardation? Does this mean that~(\ref{E_ret}) is superior, taking into account `higher order' corrections? Thinking about it now, the radial nature of a Lorentz transformed Coulomb field did seem suspicious from the start; basically too good to be true. It is as if the field would have to `know' in advance where a moving charge will be, so that it could adjust its own direction `on time' to a present charge position, so as to `point' perfectly to or from the charge. And after all, field deformation and signal retardation are complicated effects, each by itself. Surely, their combined result can only be more complicated---they certainly can not `conspire' so as produce a simpler result, a purely radial field for a uniformly moving charge. Can they?

Before proceeding to the answer (which a reader may very well already know), let us again take a moment to appreciate that the question is perfectly legitimate, especially for a student who is for the first time presented with~(\ref{E_ret}). In fact, that the field of a moving charge should still be radial---thus pointing precisely at or from the present charge position, however far an observer might be---sounds suspiciously like an infinitely fast, instantaneous transfer of information. Confusion gains even greater legitimacy in light of the so called Penrose-Terrell effect, concerning the visual appearance of a rapidly (relativistically) moving object \cite{terrell_1,terrell_2,terrell_3,terrell_4,terrell_5}. In order to describe the \textit{appearance} of the moving object, one must first take into account its Lorentz contraction---akin to a field deformation from~(\ref{E_ret})---an then, \textit{separately}, a finite time required for light rays to reach an optical device (e.g. an eye or a camera), which is precisely a retardation effect.


\begin{figure}[b!]
\centering
\includegraphics[width=0.3\textwidth,keepaspectratio]{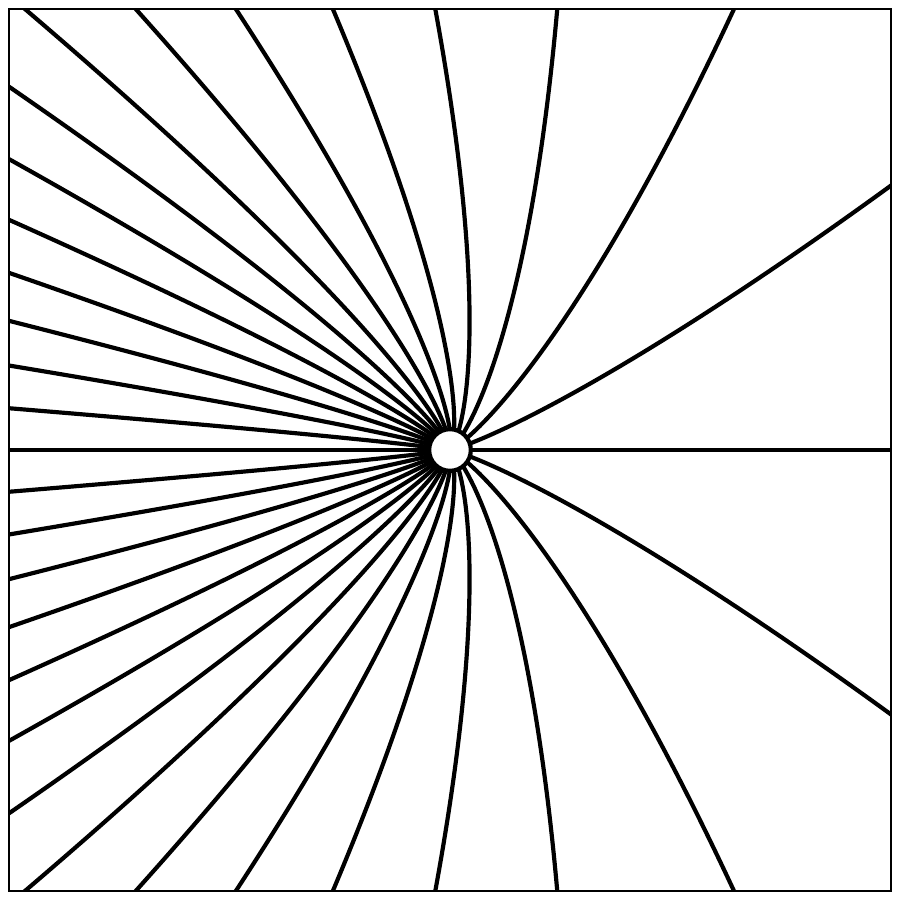}
\caption{Field lines of a point charge in a uniform relativistic motion to the right, with speed \mbox{$v=0.15c$}, \textit{as possibly imagined} by a student after being presented with the concept of signal retardation from~(\ref{E_ret}). These particular field lines actually represent a field from within the square brackets, \textit{without} the retardation effect having been taken into account.} 
\label{fig2}
\end{figure}

In this sense, should an electric field of a uniformly moving charge (as represented by field lines) actually look like the one from figure~\ref{fig2}? For a charge moving to the right, should the field lines actually be `trailing' behind it due to a signal retardation? Alas, no. In fact, figure~\ref{fig2} represents the field from \textit{within} the square brackets in~(\ref{E_ret}), \textit{without} the retardation effect taken into account (a pure field deformation, as we have defined it). In reality, these field lines more closely (but not accurately) resemble a field of a uniformly \textit{accelerated} charge~\cite{uniform_acc1,uniform_acc2,uniform_acc3}.


Equation~(\ref{E_ret}) does, in fact, reproduce a radial field from~(\ref{Er}), precisely due to a retardation effect. This is easily shown with very little calculation. To this end we need to show that the direction of a retarded quantity \mbox{$[\hat{\mathbf{R}}-\boldsymbol{\beta}]_\tau=\hat{\mathbf{R}}_\tau-\boldsymbol{\beta}$} reduces to a charge-relative radial direction~$\hat{\mathbf{R}}_t$ \textit{at the moment of field observation}\footnote{
Throughout this work we either use a notation such as~$\mathbf{R}_t$ or the notation like~$\mathbf{r}_0(t)$ to denote a time dependence of the relevant quantities. Both notations are equivalent.
}. Due to a uniform charge motion, $\boldsymbol{\beta}$ is constant in time. In defining \mbox{$\mathbf{R}= \mathbf{r}-\mathbf{r}_0$}, only a charge position~$\mathbf{r}_0$ is a function of time. Since its time dependence is known from a uniformity of motion: \mbox{$\mathbf{r}_0(t)-\mathbf{r}_0(\tau)=(t-\tau)c\boldsymbol{\beta}$}, it directly follows:
\begin{equation}
\mathbf{R}_\tau=\mathbf{R}_t+(t-\tau)c\boldsymbol{\beta}.
\label{Rtau}
\end{equation}
Using this to construct a retarded quantity \mbox{$[\hat{\mathbf{R}}-\boldsymbol{\beta}]_\tau$} from~(\ref{E_ret}), we have:
\begin{equation}
\hat{\mathbf{R}}_\tau-\boldsymbol{\beta}=\frac{\mathbf{R}_t+(t-\tau)c\boldsymbol{\beta}}{R_\tau}-\boldsymbol{\beta}.
\label{Rb_ret}
\end{equation}
On the other hand, a general condition \mbox{$\tau+R_\tau/c=t$} for a retarded time yields:
\begin{equation}
R_\tau=(t-\tau)c.
\label{ret_cond}
\end{equation}
Plugging this into~(\ref{Rb_ret}) leaves:
\begin{equation}
\hat{\mathbf{R}}_\tau-\boldsymbol{\beta}=\frac{\mathbf{R}_t}{R_\tau},
\label{Rb_radial}
\end{equation}
i.e. a retarded construction \mbox{$[\hat{\mathbf{R}}-\boldsymbol{\beta}]_\tau$} is indeed radial relative to a \textit{present} charge position: \mbox{$[\hat{\mathbf{R}}-\boldsymbol{\beta}]_\tau\propto \hat{\mathbf{R}}_t$}. A proof that the rest of~(\ref{E_ret}) reduces completely to~(\ref{Er}) may be found in Appendix~\ref{appendix_equality}.

In spite of an equivalence between~(\ref{Er}) and~(\ref{E_ret}) being \textit{almost} too good to be true, it turns out to be perfectly true nonetheless. This fact is even more remarkable once it has been demonstrated, than it ever could have been while it was just suspected. Yet, even in the face of an extraordinary nature of this outcome, one can easily understand why the retarded quantities are not explicitly present in~(\ref{Er}). A reason is that a uniform motion is \textit{perfectly specified} so that all present and retarded quantities are directly related, i.e. all retarded quantities may be expressed via present ones.

Teachers may also take note of another reason for~(\ref{E_ret}) to confuse the students, even if the students seemed to have grasped the general concept of retardation. It is the simple fact that in~(\ref{E_ret}) the mathematical effect of retardation is `hidden from the eyes'. Accounted for by a very abstract notation $[\cdot]_\tau$, its specific effect upon the rest of the expression is entirely nontransparent. What \textit{is} directly accessible by eye is the term \mbox{$\hat{\mathbf{R}}-\boldsymbol{\beta}$}. By itself this term is evidently not radial so is it any wonder that a student might imagine~(\ref{E_ret}) as describing a non-radial field, such as the one from figure~\ref{fig2}?


Even after demonstrating an equivalence between~(\ref{Er}) and~(\ref{E_ret}), their relation is not rendered trivial. As we have seen in Section~\ref{sec_radial}, the Lorentz transformation of an electric field may be derived from the coordinate transformations, where Lorentz contraction of spatial coordinates plays a central role, but the retardation effect never needs to be explicitly taken into account. On the other hand, classical derivations of~(\ref{E_general}) in which a retardation is explicit, never make an explicit reference to a Lorentz contraction. This issue is clearly pointed out and presented in detail by Jefimenko~\cite{retarded1,retarded2}, who concludes in~\cite{retarded2}: `[...] the reason why Lorentz contraction can be ignored in the classical solution [...] remains unclear.' This lends further legitimacy to the students' confusion about the relation between~(\ref{Er}) and~(\ref{E_ret}). The moral here is that such confusions should not be disregarded as banal.

\subsection*{Let us put it to the test}

\begin{figure*}[t!]
\centering
\includegraphics[width=0.32\textwidth,keepaspectratio]{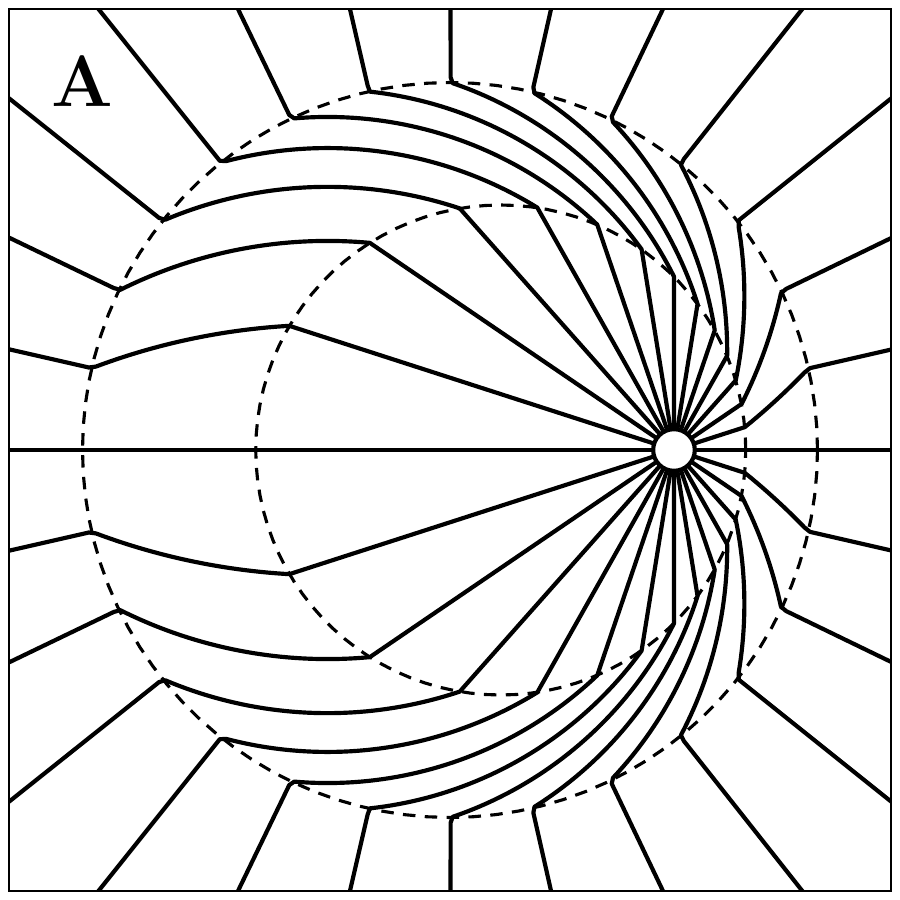}
\includegraphics[width=0.32\textwidth,keepaspectratio]{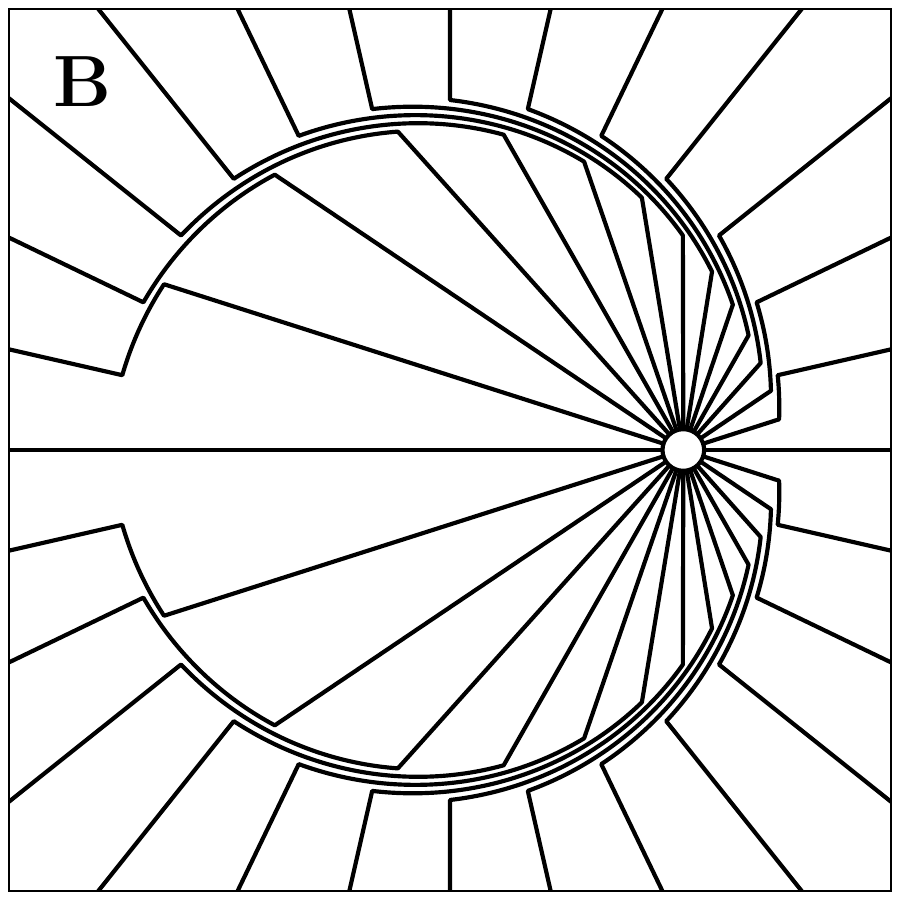}
\includegraphics[width=0.32\textwidth,keepaspectratio]{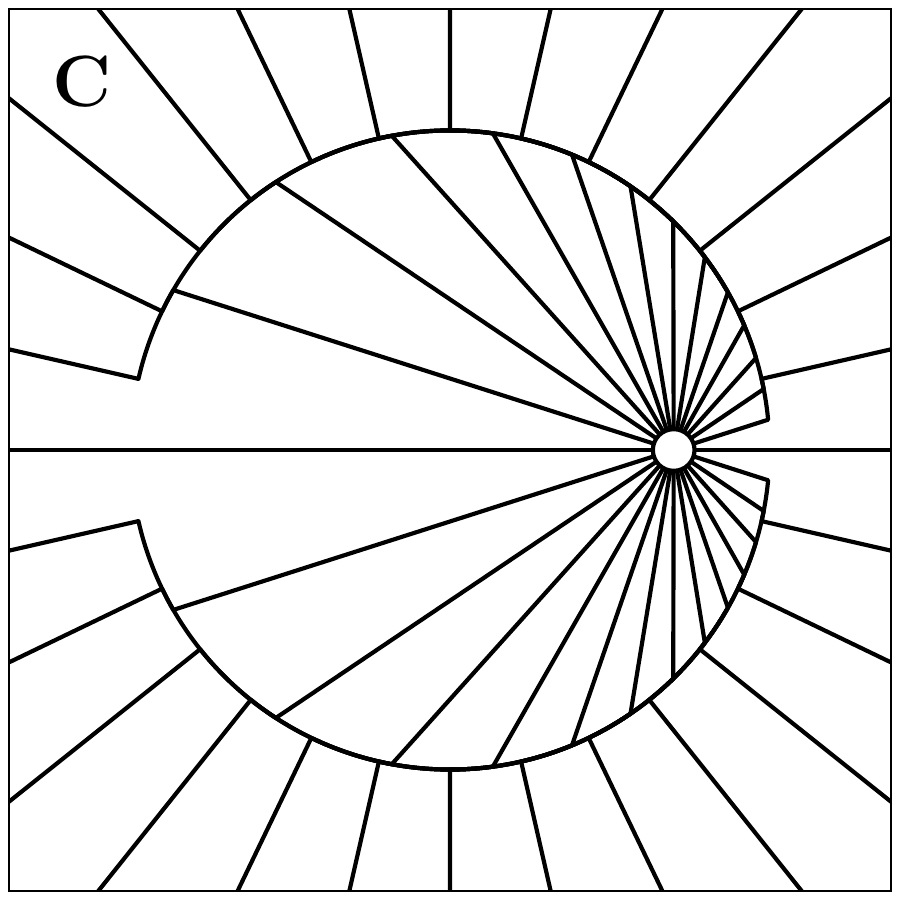}
\caption{Field line representation of an electric field of the point charge initially at rest. It is then accelerated to a uniform horizontal motion to the right, with speed \mbox{$v=0.7c$}. Frame~A shows a case of a finite acceleration. Frame~C illustrates the instantaneous acceleration. Frame~B may be viewed either as a case of a very sudden (but still finite) acceleration, or just as an artificial way of illustrating the instantaneous acceleration, so as to avoid a visual overlap of field lines from frame~C.}
\label{fig3}
\end{figure*}

There is a nice example to be used in testing the students' intuitive internalization of the retardation effect. It involves a transition between two states of uniform motion. As such, this example necessarily involves some acceleration, but the details of the acceleration itself and its effect upon the electric field will be completely irrelevant for the main point of discussion. Nevertheless, a short introductory discussion is in order to set the stage and provide a clear physical picture.

Let the point charge initially be at rest. At the moment~$t_1$ let it accelerate (uniformly or otherwise) from the initial position~$\mathbf{r}_1$. Let it keep accelerating until~$t_2$, when it reaches a point~$\mathbf{r}_2$ and continues moving with the attained, uniform velocity~$\mathbf{v}$. Let us now observe its electric field at some later moment~$t_3$, when the charge is at \mbox{$\mathbf{r}_3=\mathbf{r}_2+(t_3-t_2)\mathbf{v}$}. Prior to~$t_1$ its field was purely electrostatic. During acceleration the field changes according to~(\ref{E_general}), but \textit{only} within a limited portion of space that the information about the change in charge motion can reach in a given time. This is, of course, a spherical interior centered at~$\mathbf{r}_1$, with a radius of~$c(t-t_1)$, $t$~being a general moment of field observation. After attaining a final velocity at~$t_2$ and continuing with a uniform motion, the field around the charge is a familiar Coulomb field. But again, only within a portion of space than the information about this new uniform motion can reach. This is now a spherical interior centered at~$\mathbf{r}_2$, with the radius of \mbox{$c(t_3-t_2)$}, $t_3$ being a selected moment of the field observation, when the charge is at~$\mathbf{r}_3$. Such electric field is illustrated by the first (A) frame from figure~\ref{fig3}. In particular, it shows the field lines of a point charge  at some moment~$t_3$, moving horizontally to the right with a speed of \mbox{$v=0.7c$}, the acceleration having being uniform (in a relativistic sense~\cite{rel_acc}) between~$t_1$ and~$t_2$. Two described spherical boundaries are indicated by the dashed lines. Outer one corresponds to the information boundary related to the start of acceleration at~$t_1$; inner one to the end of acceleration at~$t_2$. Field line sections in between these boundaries describe a field affected by the charge acceleration. This example is well known and well investigated~\cite{finite_acc1,finite_acc2}.


After giving this physically sound motivation, a subsequent discussion is greatly simplified by shifting a focus away from acceleration. This is easily achieved by considering an \textit{instantaneous} charge acceleration (in a sense of a limiting process \mbox{$t_2\to t_1$}) from a rest state to a uniform motion with a given velocity; as if a charge was instantly `launched' in a bout of infinite acceleration of infinitesimal duration. This case is illustrated by the third~(C) frame from figure~\ref{fig3}. However physically artificial, this scenario \textit{is} admitted by Maxwell equations. Since the acceleration is instantaneous, it may be described by a Dirac delta-function: \mbox{$\mathbf{a}=\mathbf{v}\delta(t-t_1)$}, which then appears in the electric field through the acceleration term from~(\ref{E_general}). This singular contribution to the field is responsible for a sudden `break' in the field lines from frame~C. Students interested in further technical details can always be referred to a relevant literature~\cite{delta}. Admittedly, there is a small visual drawback to a display from frame~C, since some sections of field lines along the transition sphere are now overlapped. This makes it difficult to visually connect their `inner' and `outer' sections. For this reason the sources discussing an instantaneous~\cite{delta} or a very sudden~\cite{sudden_acc1} acceleration usually display the corresponding field lines in a manner shown by the second (B) frame from figure~\ref{fig3}. Evidently, it better preserves the visual identity of individual field lines, but in our opinion heavily detracts from the two states of uniform motion. We leave it to the reader to decide which is a lesser price to pay.

\begin{figure*}[t!]
\centering
\includegraphics[width=0.32\textwidth,keepaspectratio]{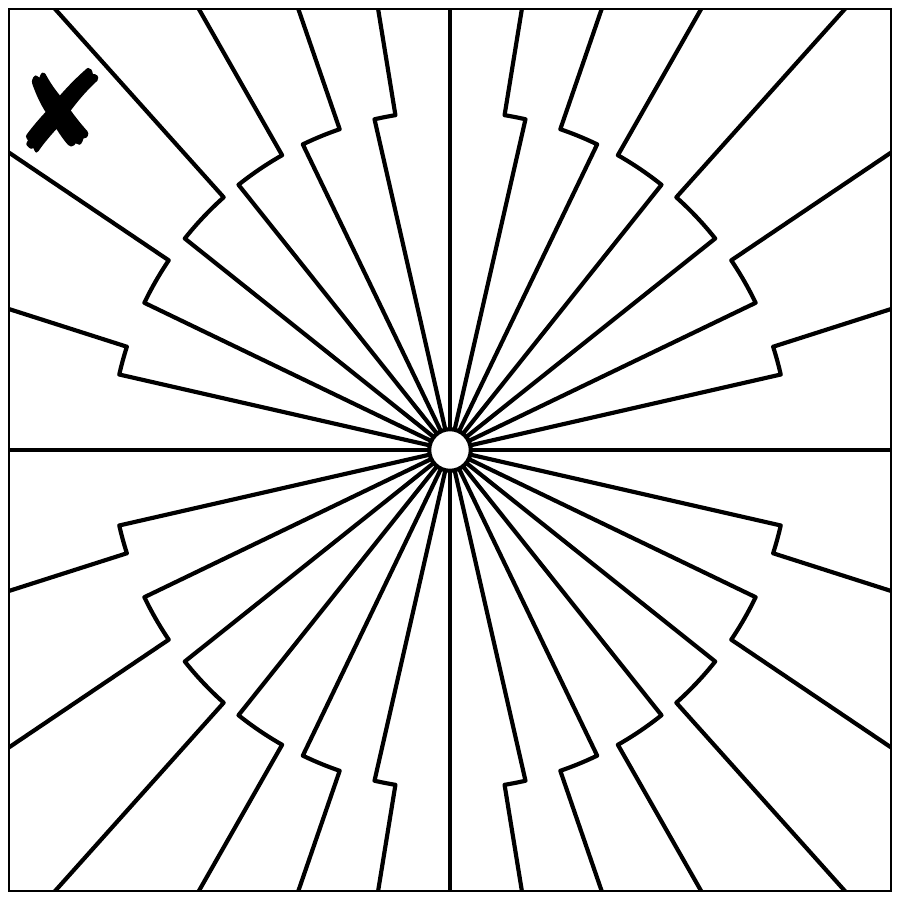}
\includegraphics[width=0.32\textwidth,keepaspectratio]{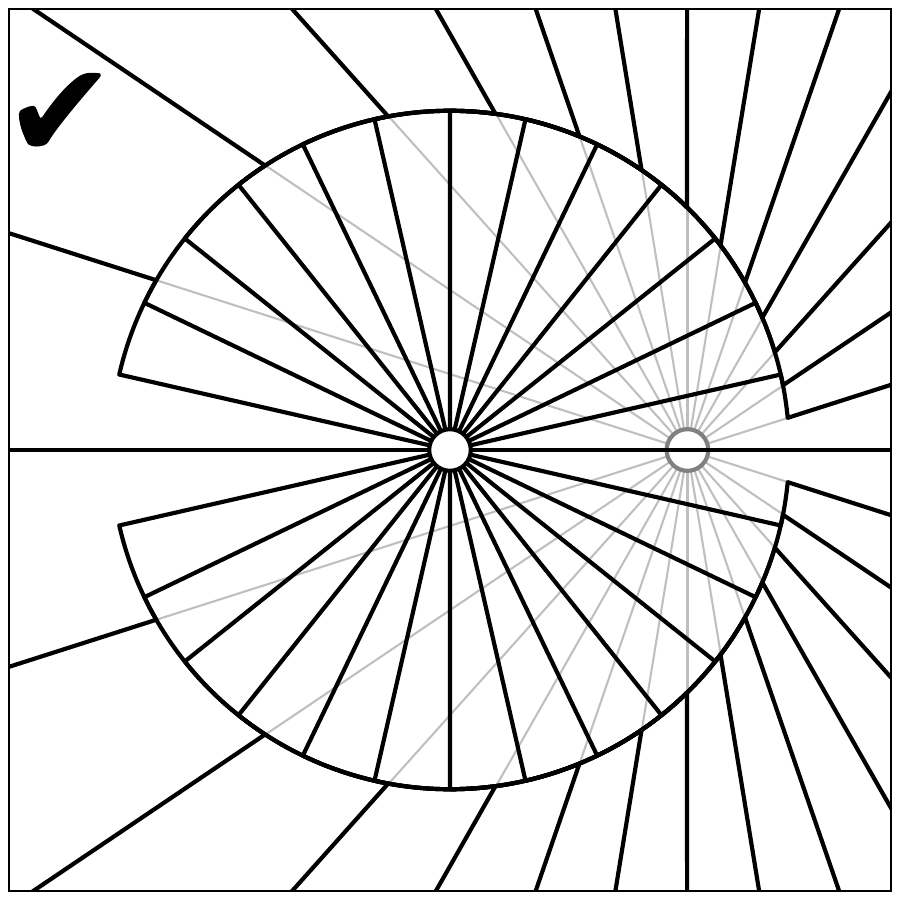}
\caption{Field lines of a point charge instantaneously stopped from a horizontal uniform motion to the right, with \mbox{$v=0.7c$}. Left plot (\ding{56}) is an intuitive but incorrect depiction. Right plot (\ding{52}) is a correct one.
} 
\label{fig4}
\end{figure*}

Now that the physical introduction has been made, we arrive at the central question that you may want to ask your students. Simply ask them to draw the field lines of a point charge instantaneously \textit{stopped} from a uniform motion. There is a good chance you may get an answer from the left plot in figure~\ref{fig4}. It attempts to show a charge instantaneously stopped from \mbox{$v=0.7c$}, in a manner analogous to a frame~C from figure~\ref{fig3}. This answer is marked with~\ding{56} for a simple reason that it is wrong. Certainly, the sections of field lines inside the transition sphere are isotropic, indicating an electrostatic field, while the outer sections are Lorentz contracted, indicating a past uniform motion. Thus the retardation effect has been taken into account. But not all of it! For the outside sections of field lines to remain centered at the resting charge, the field lines (i.e. an electric field they illustrate) would have to `know' that the charge has stopped. But how could they know it if the information about the charge having stopped has not yet reached this portion of space; has not had a chance to propagate outside the transition sphere? Now is a crucial moment for students to realize---if they have not already---how deep the retardation effect goes. A field outside the transition sphere must keep evolving as if the charge was still in motion, and the field lines must keep `moving' correspondingly. Thus, the correct depiction of the field lines is shown by the right plot from figure~\ref{fig4}, the one marked with~\ding{52}. Lighter-in-color extensions of the outer segments converge at the position where the charge would have been at the moment of field observation if it kept moving with a uniform speed \mbox{$v=0.7c$}. This behavior is self-evident when using~(\ref{E_general}) to calculate a field of a charge subject to \textit{finite} deceleration, thus obtaining `non-pathological' field lines. Still, initial intuitive appeal of the incorrect answer from figure~\ref{fig4} should be recognized and appreciated. Not only does the correct continuation of the field line `motion' outside the transition sphere seem to be unexpected among students when they are asked to produce the plot by themselves. Rather, if they are presented from the start by a correct diagram (avoiding a discussion of an incorrect one), they often fail to register anything significant about the straight field line segments that have `run away' from a stopped charge. That it \textit{is} significant, in fact counterintuitive, is revealed after their attention is brought to this detail.


In the end, students often find it rewarding to realize, either by themselves of with a teacher's guidance, that the transient portions of field lines from figure~\ref{fig3} correspond to a classical description of an \textit{electromagnetic shockwave}, and those from a correct frame in figure~\ref{fig4} to a classical depiction of \textit{bremsstrahlung}.

\section{Back to the field lines}
\label{field_lines}

\begin{figure*}[t!]
\centering
\includegraphics[width=0.32\textwidth,keepaspectratio]{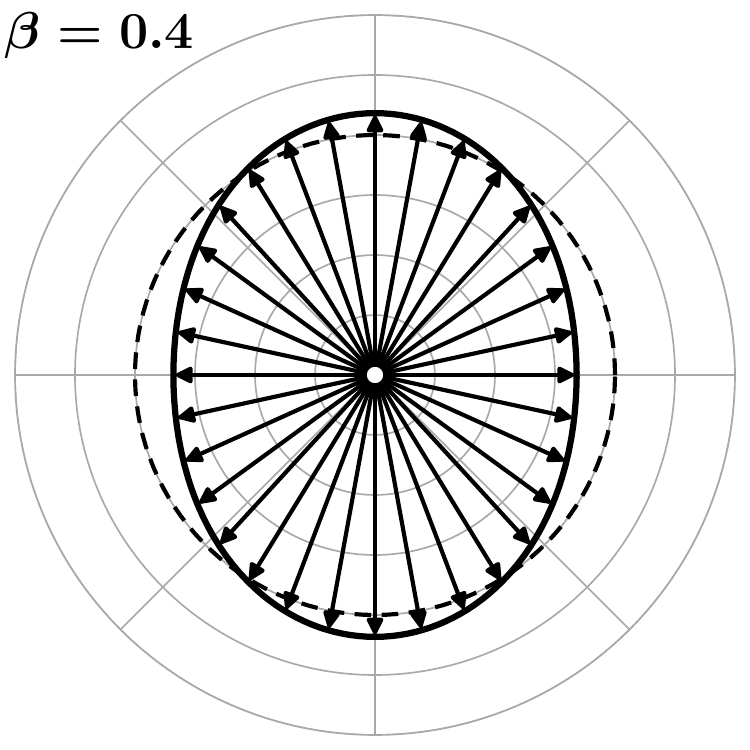}
\includegraphics[width=0.32\textwidth,keepaspectratio]{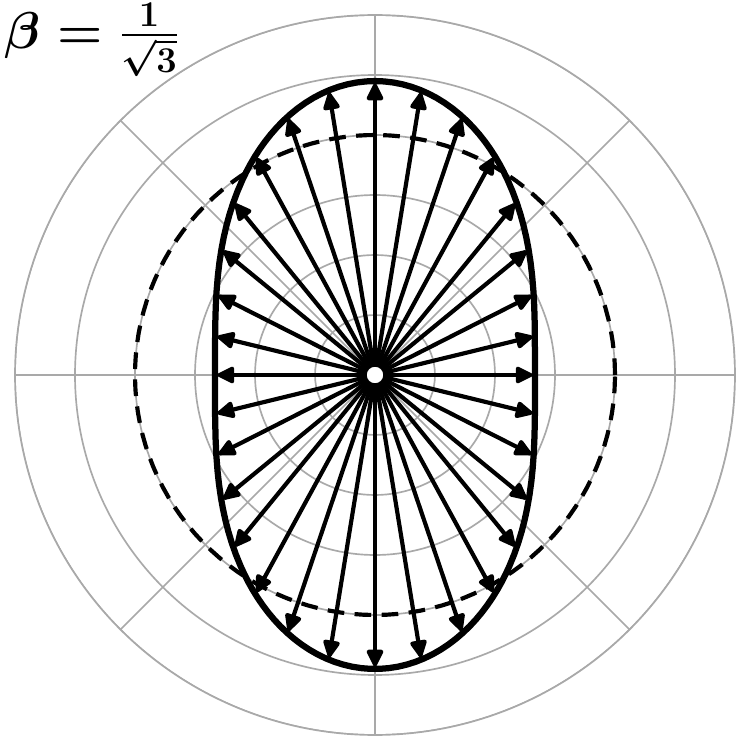}
\includegraphics[width=0.32\textwidth,keepaspectratio]{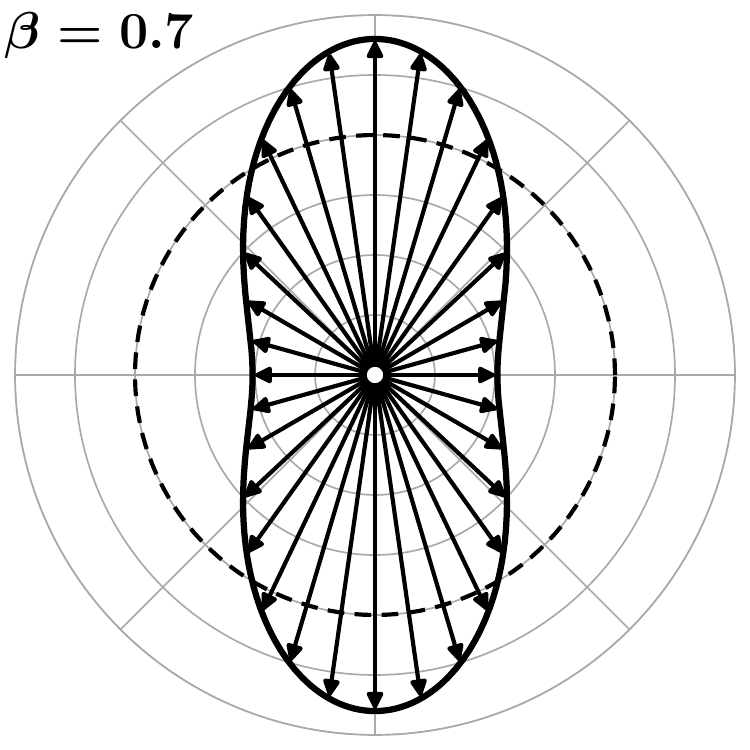}
\caption{Polar plots of the field strength from~(\ref{Er})---i.e. of the relevant part \mbox{$(1-\beta^2)/[1-\beta^2\sin^2\theta']^{3/2}$}, for different values of~$\beta$---at a fixed distance from a point charge in a uniform motion. For reference, dashed lines show a strength of an electrostatic field (\mbox{$\beta=0$}). The displayed length of superimposed field lines is proportional to a field strength.
} 
\label{fig5}
\vspace*{-1mm}
\end{figure*}

Now we may finally return to a depiction of field lines from figure~\ref{fig1}. Such depictions are primarily meant to convey a Lorentz contraction of a Coulomb field of a point charge. This is achieved by compressing the field lines of a moving charge toward the axis perpendicular to its motion. The rule for drawing the compressed field lines is simple. In a rest frame of the charge the isotropic nature of an electrostatic field is conveyed by isotropically distributed field lines. Let~$\theta$ be an angle of one such particular field line (in a rest frame) relative to a direction of charge motion (from a frame where it moves). In a frame where it moves with speed~$v$, the `same' field line is to be drawn under an angle $\theta'$ such that:
\begin{equation}
\tan\theta'=\gamma\tan\theta.
\label{tans}
\end{equation}
Feynman gives a beautify simple argument for this~\cite{feynman}: just perform a Lorentz contraction of uniformly distributed field lines from a charge rest frame (as if the field lines were real physical objects, subject to coordinate transformations). Another argument is a bit more complex and is based on the Gauss's law~\cite{purcell,finite_acc1}. It considers a `conservation' of the electric flux for a charge accelerated between two states of uniform motion (such as in figures~\ref{fig3} and~\ref{fig4}), wherein the appropriate angle for a particular field line is determined from a requirement that the entire field line lies along certain (quasi)conical boundaries circumscribing a region of a conserved electric flux.

What needs to be understood is that the field lines are mathematical artifacts, integral curves that have no physical meaning or consequence, nor are subject to any physical law. From the information point of view they are entirely redundant, since they carry no additional information about the corresponding vector field that is not already contained within the field itself~\cite{redundant}. Therefore, there is no \textit{a priori} reason to draw the field lines of a moving charge in any relation to the field lines of a charge at rest. It would be perfectly correct to draw them isotropically or distributed in any other way. How we draw them critically depends on the message we are trying to convey.

So what are the different depictions from figure~\ref{fig1} trying to convey? Let us start with a depiction~B, encountered as figure~5.15 from~\cite{purcell}. This seems to be the most popular depiction found in literature, and for a good reason. The compression of field lines conveys a message about the Lorentz contraction of coordinates in a direction of a charge motion, and nothing else. It does so in a manner which is not redundant and thus not open to misinterpretation. In particular, the length of the field lines is selected to be the same as for the charge at rest.

Depiction~C, encountered as figure~26-4 from~\cite{feynman}, tries to go a step further. It aims to additionally emphasize the Lorentz contraction by shortening the field lines length in a direction of charge motion. This redundancy would not be questionable---would even be commendable for clarity---if it were not susceptible to a gross misinterpretation. A danger lies in trying to interpret the field line lengths as depicting an electric field strength in a given direction (under a tacit assumption of a fixed distance from a point charge). Even for an attentive reader who carefully reads the description of such drawing, this may become an issue later if only the visual appearance of a drawing remains committed to memory, without a full and meaningful description of it.

As seen from~(\ref{E_para}) and~(\ref{E_perp}), a forward $E_x$ component is reduced by a factor~$1/\gamma^2$, while a lateral $E_y$ component is increased by a factor~$\gamma$, relative to the components of an electrostatic field (that would be present if, \textit{in the same frame}, the charge was at rest instead of being in motion). Depiction~C, on the other hand, shows an unmodified length of a lateral field line. At the same time, a length of a forward of backward field line is reduced only by a factor $1/\gamma$, coming from a Lorentz length contraction. Thus, not even a relative ratio $1/\gamma^3$ of these field components is intended to be conveyed by this depiction. Secondly, by free association a viewer may attempt to use a memory of depiction~C in recalling the Lorentz transformations from~(\ref{lor_ex}) and~(\ref{lor_ey}), especially since a relative difference between thus transformed components is precisely $1/\gamma$. In this way a viewer might mistakenly conclude that $E_y$ component remains the same while $E_x$ component decreases, in a complete opposition with correct transformations.


The third (unintended) issue with depiction~C is the most subtle one. Since the length of the shortened field lines is supposed to reflect the length contraction, the outline they form is elliptical in shape. By (falsely) associating this length with the field strength from~(\ref{Er}), one might imagine that the relevant part \mbox{$(1-\beta^2)/[1-\beta^2\sin^2\theta]^{3/2}$} of this equation describes an ellipse (we again omit a primed notation). That this is false is a trivial matter. What is interesting is \textit{how} false it is. Due to a forward-backward symmetry of~(\ref{Er}), an angular domain relevant for a discussion is \mbox{$\theta\in[0,\pi/2]$}. Within this range the norm \mbox{$E(\theta)$} of~(\ref{Er}) is evidently monotonous in~$\theta$. The same holds for $E_y$ component \mbox{$E_y(\theta)=E(\theta)\sin\theta$}. However, $E_x$ component \mbox{$E_x(\theta)=E(\theta)\cos\theta$} is not always monotonous! This does not seem to be a well addressed issue in the literature so we address it here. By a simple maximization procedure one finds that $E_x(\theta)$ is supposed to have a maximum at:
\begin{equation}
\theta_\pm=\acos\left(\pm\frac{1}{\beta\gamma\sqrt{2}}\right)=\acos\left[\pm\sqrt{\frac{1}{2}\left(\frac{c^2}{v^2}-1\right)}\right],
\label{theta_pm}
\end{equation}
where $\theta_+$ corresponds to a forward angle and $\theta_-$ to a backward one: \mbox{$\theta_-=\pi-\theta_+$}. However, the argument under the arcus cosine function can be at most 1, which is satisfied only for \textit{$v>c/\sqrt{3}$}. Therefore, only under this condition does $E_x$ component have \textit{nontrivial} maxima. The maximizing values may be parametrized as:
\begin{equation}
\theta_{\max}=\left\{\begin{array}{lcc}
0 \;\mathrm{or}\; \pi&\mathrm{if}& \beta\le 1/\sqrt{3},\\
\theta_+ \;\mathrm{or}\; \theta_-&\mathrm{if}& \beta> 1/\sqrt{3},
\end{array}\right.
\end{equation}
while the maximum $E_x$ component itself turns out to be:
\begin{equation}
\Big|\Big(\frac{q}{4\pi\epsilon_0 R^2}\Big)^{-1} E_x\Big|_{\max}=
\left\{\begin{array}{lcc}
1/\gamma^2&\mathrm{if}& \beta\le 1/\sqrt{3},\\
2/(3\sqrt{3}\beta)&\mathrm{if}& \beta> 1/\sqrt{3}.
\end{array}\right.
\label{Ex_max}
\end{equation}
At the `phase transition' value \mbox{$\beta=1/\sqrt{3}$} both cases reduce to 2/3. We see that \textit{only} for \mbox{$v\le c/\sqrt{3}$} does the commonly quoted factor $1/\gamma^2$ from~(\ref{E_para}) represent the upper bound for the $E_x$ component (under a tacit assumption of a fixed distance from a moving charge). The novelty here is that for \mbox{$v\to c$}, i.e. \mbox{$\beta\to 1$} the $E_x$ component does not entirely vanish! In fact, under $\theta_{\max}$ the numerical factor from~(\ref{Ex_max}) saturates at \mbox{$2/(3\sqrt{3})\approx0.385$}. Thus, even in the ultrarelativistic case, $E_x$ component may be as high as 38.5\% of a `corresponding electrostatic' field strength! Commonly quoted suppression of $E_x$ component by $1/\gamma^2$ strictly refers to a field right ahead or behind a moving charge (for \mbox{$\theta=0$} or $\pi$).

Figure~\ref{fig5} shows polar plots illustrating the $E(\theta)$ dependence for values of $\beta$ below, precisely at and above the critical value of $1/\sqrt{3}$, at some fixed distance from a moving charge (a polar plot of the field strength should not be confused with a geometrical distance from the charge). In each of them a dashed circle represents an isotropic electrostatic field, as a point of reference. For \mbox{$\beta<1/\sqrt{3}$} the field strength displays a commonly imagined dependence, roughly resembling an ellipse. For \mbox{$\beta>/\sqrt{3}$} the shape of the angular field dependence becomes far more interesting, featuring a maximum $E_x$ component at nontrivial angles from~(\ref{theta_pm}).

Based on these observations we propose drawing the field lines from figure~\ref{fig1} in a manner represented by a frame~A, wherein the length of the drawn lines reflects an angular dependence $E(\theta)$ of the field strength. The reason is simple: radial field lines \textit{do} allow this additional `degree of freedom' to be represented in this way. In comparison with frame~C, a far more important reason is that this display avoids a danger of misinterpreting the field line lengths, by the virtue of them showing precisely what one would expect. While the display from frame~B is still satisfactory in this manner (free from misinterpretation)---and very much formally correct---it feels like a `missed opportunity' for conveying additional interesting and relevant information. To further illustrate our proposal, all plots from figure~\ref{fig5} have been complemented with sets of such field lines (it should be noted that these field lines `live' in a geometric space, while the field strength plots `live' in a vector space spanned by the electric field components). Frame~A from figure~\ref{fig1} precisely corresponds to a \mbox{$\beta=0.7$} case from figure~\ref{fig5}, which finally explains the unfamiliar field line lengths that a reader might have been wondering about.

\section{Conclusions}

Due to a recent direct experimental confirmation of a Lorentz contraction of a Coulomb field~\cite{nature}, we expect a second renaissance in teaching the relativistic transformations of electric and magnetic fields. For this reason we have revisited some widespread confusions about the electric field of a point charge in uniform motion. We believe that every teacher of this topic should be aware of these confusions, in order to address them in a timely manner. Our main goal here was to present how these confusions might appear in the first place and to give a centralized overview of otherwise well known solutions, which may usually be found scattered throughout the literature. We do hope that we have also provided some novel insights by approaching a subject from multiple angles.

Main confusions regarding an electric field of a point charge in uniform motion are related to its radial nature (is it really radial and how can it be so) and to a role of a signal retardation effect (is it already accounted by the common relativistic expressions encountered in literature or should it be added on top of them). We have shown---and shown how to convincingly show---that `all is right with the world:' the field in question is indeed radial and the retardation effect is indeed contained within the familiar equations obtained by Lorentz transformations.

We have also addressed a common practice of depicting a Lorentz contracted field by means of Lorentz contracted field lines. We have reason to believe that we have provided some new novel insights on this topic, by proposing an improved way for depicting them. Common field line depictions encountered in literature only illustrate the Lorentz contraction of spatial coordinates, not showing or even suggesting a false transformation of the electric field strength. These field line diagrams do have a `degree of freedom' that can be used to convey this additional information. If used properly, it also removes a danger of misinterpreting alternative depictions.

\section*{Acknowledgments}

The work of I.S. was supported by the Croatian Science Foundation Project No. IP-2020-02-9614.

\appendix

\section{Length contraction justification}
\label{appendix_contraction}

We address here a subtlety in a detailed derivation of the length contraction \mbox{$R_x'=R_x/\gamma$} from~(\ref{Rxy_old}) and ~(\ref{Rxy_new}). In order to observe the \textit{simultaneous} distance between the points $\mathbf{r}'$ and $\mathbf{r}'_0$ in a target frame (where the charge is in motion):
\begin{equation}
t'(\mathbf{r}')-t'(\mathbf{r}'_0)=\gamma \big[t(\mathbf{r}) + vx/c^2\big]-\gamma\big[t(\mathbf{r}_0) + vx_0/c^2\big]=0,
\end{equation}
the corresponding points $\mathbf{r}$ and $\mathbf{r}_0$ from the source frame (where the charge is at rest) must have bin asynchronous:
\begin{equation}
\delta t = t(\mathbf{r}) - t(\mathbf{r}_0) = - v(x-x_0)/c^2 \neq 0,
\end{equation}
with $x$ and $x_0$ as $x$-components of the position vectors (\mbox{$x=\mathbf{r}\cdot\hat{\mathbf{x}}$} and \mbox{$x_0=\mathbf{r}_0\cdot\hat{\mathbf{x}}$}). Therefore, the spatial distances $R_x'$ and $R_y'$ between simultaneous points in the target frame---used in~(\ref{Rxy_new}) for expressing the radial nature of the field---correspond to spatial distances $R_x$ and $R_y$ between \textit{asynchronous} points in the source frame. Writing out~(\ref{Rxy_old}) in explicit detail:
\begin{equation}
\frac{E_x'\big[\mathbf{r}'(t')\big]}{E_y'\big[\mathbf{r}'(t')\big]}=
\frac{1}{\gamma}\frac{E_x[\mathbf{r}(t)\big]}{E_y[\mathbf{r}(t)\big]}=
\frac{1}{\gamma}\frac{\big[\mathbf{r}(t)-\mathbf{r}_0(t+\delta t)\big]\cdot \hat{\mathbf{x}}}{\big[\mathbf{r}(t)-\mathbf{r}_0(t+\delta t)\big]\cdot \hat{\mathbf{y}}},
\label{app_A3}
\end{equation}
at first glance suggests that the application of equality $E_x/E_y=R_x/R_y$ within~(\ref{Rxy_old}) is false and unjustified because in~(\ref{radial}) the spatial distances $R_x$ and $R_y$ are between the points \textit{simultaneous in a source frame}:
\begin{equation}
\frac{E_x[\mathbf{r}(t)\big]}{E_y[\mathbf{r}(t)\big]}=\frac{\big[\mathbf{r}(t)-\mathbf{r}_0(t)\big]\cdot \hat{\mathbf{x}}}{\big[\mathbf{r}(t)-\mathbf{r}_0(t)\big]\cdot \hat{\mathbf{y}}}.
\end{equation}
However, the application of the stated equality \textit{is} valid after all because---and only because---a charge in a source frame is at rest. Its position $\mathbf{r}_0$ is constant, i.e. it holds \mbox{$\mathbf{r}_0(t)=\mathbf{r}_0(t+\delta t)$} for any~$\delta t$, so that in (\ref{app_A3}): \mbox{$\mathbf{r}(t)-\mathbf{r}_0(t+\delta t)=\mathbf{r}(t)-\mathbf{r}_0(t)$}.

\section{Equality between equations~(\ref{Er}) and~(\ref{E_ret})}
\label{appendix_equality}

The easiest way to show that~(\ref{E_ret}) reduces completely to~(\ref{Er}) is to posit their equality:
\begin{equation}
\frac{q}{4\pi\epsilon_0\gamma^2}\bigg[\frac{\hat{\mathbf{R}}-\boldsymbol{\beta}}{K^3R^2}\bigg]_\tau=\frac{q}{4\pi\epsilon_0\gamma^2}\frac{1}{[1-\beta^2\sin^2\theta_t]^{3/2}}\frac{\mathbf{R}_t}{R_t^3}
\label{eqA}
\end{equation}
and prove it to be true. Here we have immediately adapted the notation from~(\ref{Er}) to the requirements of this derivation. 
Using~(\ref{Rb_radial}), the relevant part of~(\ref{E_ret}) reduces to:
\begin{equation}
\bigg[\frac{\hat{\mathbf{R}}-\boldsymbol{\beta}}{K^3R^2}\bigg]_\tau =\frac{\mathbf{R}_t}{K_\tau^3R_\tau^3}.
\label{eqB}
\end{equation}
Rewriting~(\ref{Rb_radial}) in a form:
\begin{equation}
\mathbf{R}_\tau=\mathbf{R}_t+R_\tau\boldsymbol{\beta}
\label{eqC}
\end{equation}
and using a definition \mbox{$K_\tau=1-\hat{\mathbf{R}}_\tau\cdot\boldsymbol{\beta}$} yields:
\begin{equation}
K_\tau R_\tau=(1-\beta^2)R_\tau-\mathbf{R}_t\cdot\boldsymbol{\beta}.
\label{eqD}
\end{equation}
Plugging (\ref{eqD}) into (\ref{eqB}) and then into (\ref{eqA}) leaves:
\begin{equation}
[(1-\beta^2)R_\tau-\mathbf{R}_t\cdot\boldsymbol{\beta}]^3=[1-\beta^2\sin^2\theta_t]^{3/2}R_t^3
\end{equation}
to be proven. Using \mbox{$\mathbf{R}_t\cdot\boldsymbol{\beta}=\beta R_t\cos\theta_t$} and raising the previous equation to the power of 2/3 gives:
\begin{equation}
[(1-\beta^2)R_\tau-\beta R_t\cos\theta_t]^2=(1-\beta^2\sin^2\theta_t)R_t^2.
\end{equation}
After some algebra, the following remains:
\begin{equation}
(1-\beta^2)R_\tau^2-2\beta R_\tau R_t \cos\theta_t -R_t^2=0
\end{equation}
as a prerequisite for equality from~(\ref{eqA}) to be true. This result is confirmed by squaring in a dot product sense a \textit{known} relation from~(\ref{eqC}): \mbox{$\mathbf{R}_\tau\cdot\mathbf{R}_\tau=(\mathbf{R}_t+R_\tau\boldsymbol{\beta})\cdot(\mathbf{R}_t+R_\tau\boldsymbol{\beta})$}. Therefore,~(\ref{eqA}) holds true.


\begin{thebibliography}{99}

\bibitem{nature} Masato Ota et al., `Ultrafast visualization of an electric field under the Lorentz transformation,' Nat. Phys. \textbf{18}, 1436--1440 (2022).

\bibitem{feynman} Richard P. Feynman, Robert B. Leighton, Matthew Sands, \textit{The Feynman Lectures on Physics, Vol. II; The New Millennium Edition: Mainly Electromagnetism and Matter} (Basic books, New York, 2013), Chapter~26.

\bibitem{griffiths} David J. Griffiths, \textit{Introduction to electrodynamics}, 4th edition (Pearson Education, 2013), Chapter 10.3.

\bibitem{purcell} Edward M. Purcell, David J. Morin, \textit{Electricity and magnetism}, 3rd edition (Cambridge University Press, New York, 2013), Chapter~5.

\bibitem{einstein} Albert Einstein, `Zur Elektrodynamik bewegter K\"{o}rper,' Ann. Phys. \textbf{17}, 891--921 (1905).

\bibitem{jefimenko_direct} Oleg D. Jefimenko, `Direct calculation of the electric and magnetic fields of an electric point charge moving with constant velocity,' Am. J. Phys. \textbf{62}, 79--85 (1994). 

\bibitem{field1} David J. Griffiths, Mark A. Heald, `Time-dependent generalizations of the Biot-Savart and Coulomb laws,' Am. J. Phys. \textbf{59}, 111--117 (1991). 
\bibitem{field2} Hamsa Padmanabhan, `A simple derivation of the electromagnetic field of an arbitrarily moving charge,' Am. J. Phys. \textbf{77}, 151--155 (2009). 
\bibitem{field3} Ashok K. Singal, `A first principles derivation of the electromagnetic fields of a point charge in arbitrary motion,' Am. J. Phys. \textbf{79}, 1036--1041 (2011).

\bibitem{terrell_1} Roger Penrose `The apparent shape of a relativistically moving sphere,' Math. Proc. Camb. Philos. Soc. \textbf{55}, 137--139 (1959).
\bibitem{terrell_2} James Terrell, `Invisibility of the Lorentz Contraction,' Phys. Rev. \textbf{116}, 1041--1045 (1959).
\bibitem{terrell_3} Victor F. Weisskopf, `The visual appearance of rapidly moving objects,' Phys. Today \textbf{13}, 24--27 (1960).
\bibitem{terrell_4} Roy Weinstein, `Observation of Length by a Single Observer,' Am. J. Phys. \textbf{28}, 607--610 (1960).
\bibitem{terrell_5} Kevin G. Suffern, `The apparent shape of a rapidly moving sphere,' Am. J. Phys. \textbf{56}, 729--733 (1987).

\bibitem{uniform_acc1} E. Eriksen, \O. Gr\o n, `Electrodynamics of Hyperbolically Accelerated Charges: I. The Electromagnetic Field of a Charged Particle with Hyperbolic Motion,' Ann. Phys. \textbf{286}, 320--342 (2000). 
\bibitem{uniform_acc2} Joel Franklin, David J. Griffiths, `The fields of a charged particle in hyperbolic motion,' Am. J. Phys. \textbf{82}, 755--763 (2014). 
\bibitem{uniform_acc3} Y. Hadad, E. Cohen, I. Kaminer, A. C. Elitzur, `Covariant electromagnetic field lines,' J. Phys. Conf. Ser. \textbf{880}, 012052 (2017). 

\bibitem{retarded1} Oleg D. Jefimenko, `Retardation and relativity: Derivation of Lorentz–Einstein transformations from retarded integrals for electric and magnetic fields,' Am. J. Phys. \textbf{63}, 267--272 (1995). 
\bibitem{retarded2} Oleg D. Jefimenko, `Retardation and relativity: The case of a moving line charge,' Am. J. Phys. \textbf{63}, 454--459 (1995). 

\bibitem{rel_acc} Edward A. Desloge, R. J. Philpott, `Uniformly accelerated reference frames in special relativity,' Am. J. Phys. \textbf{55}, 252--261 (1987).

\bibitem{finite_acc1} Jack R. Tessman, Joseph T. Finnell, `Electric Field of an Accelerating Charge,' Am. J. Phys. \textbf{35}, 523527--10 (1967).
\bibitem{finite_acc2} Daja Ruhlandt, Steffen M\"{u}hle,  J\"{o}rg Enderlein, `Electric field lines of relativistically moving point charges,' Am. J. Phys. \textbf{88}, 5--10 (2020).

\bibitem{delta} Boris M. Bolotovski\u{\i},  Alexander V. Serov, `On the force line representation of radiation fields,' Phys.-Usp. \textbf{40}, 1055--1059 (1997).

\bibitem{sudden_acc1} Roger Y. Tsien, `Pictures of Dynamic Electric Fields,' Am. J. Phys. \textbf{40}, 46--56 (1972).

\bibitem{redundant} Joseph Slepian, `Lines of Force in Electric and Magnetic Fields,' Am. J. Phys. \textbf{19}, 87--90 (1951). 

\end{thebibliography}
\end{document}